\documentclass[preprint,12pt]{elsarticle}
\journal{Nuclear Physics A}
\begin{document}
\begin{frontmatter}

\title{Improved limits on $\beta^+$EC and ECEC processes in $^{74}$Se}
\author[ITEP]{A.S.~Barabash\corref{cor1}}
\ead{barabash@itep.ru}

\author[JINR]{V.B.~Brudanin}

\author[JINR]{A.A.~Klimenko}

\author[ITEP]{S.I.~Konovalov}

\author[JINR]{A.V.~Rakhimov}

\author[UTEF]{E.N.~Rukhadze}

\author[JINR]{N.I.~Rukhadze}

\author[JINR]{Yu.A.~Shitov }

\author[UTEF]{I.~Stekl}

\author[LSM]{G.~Warot}

\author[ITEP]{V.I.~Umatov}

\cortext[cor1]{Corresponding author}

\address[ITEP]{NRC "Kurchatov Institute", Institute of Theoretical and Experimental Physics, B.\
Cheremushkinskaya 25, 117218, Moscow, Russian Federation}

\address[JINR]{Joint Institute for Nuclear Research, 141980, Dubna, Moscow region, Russia}

\address[UTEF]{Institute of Experimental and Applied Physics, Czech Technical University in Prague, 12800, Prague, Czech Republic}

\address[LSM]{Laboratoire Souterrain de Modane, F-73500, Modane, France}

\begin{abstract}
New limits on $\beta^+$EC and ECEC processes in $^{74}$Se have been obtained using 
a 600 cm$^3$ HPGe detector and an external source consisting of 1600 g of a natural selenium powder. For different $\beta^+$EC and ECEC transitions (to the ground and excited states) obtained limits are on the level $\sim (0.2-4.8)\times10^{19}$ yr at  90\% C.L.
In particular, for the potentially resonant transition into the
1204.2 keV excited state of $^{74}$Ge a lower half-life limit of $1.1\times10^{19}$ yr at 90\% C.L. has been obtained. Possibility to increase the sensitivity of such measurements is discussed. 
\end{abstract}


\end{frontmatter}

\newpage

\section{Introduction}

Search for neutrinoless double beta decay is one of the most interesting tasks in nuclear physics, particle physics and astrophysics. The discovery of this process will automatically lead to two fundamental conclusions - 1) the lepton number is violated, and 2) the neutrino is a Majorana particle. In addition, it will provide information on such fundamental problems as the absolute neutrino mass scale, the type of hierarchy and the CP violation in the lepton sector (see discussions in \cite{PAS06,BIL15a,VER16}).


The current most stringent half-life limits on $0\nu\beta\beta$ decay are of the order of $10^{25} -10^{26}$ yr (see recent reviews \cite{BAR18,BAR19,DOL19}). The standard process ($2\nu\beta\beta$), which implies also the emission of two electron antineutrinos, is the rarest nuclear decay ever observed and has been registered in ten nuclei with half-lives in the range of $10^{18} - 10^{24}$ yr (see \cite{BAR15} and references therein).

Much less attention is paid to the study of $2\beta^+$, $\beta^+$EC and ECEC
processes although such studies are constantly being conducted and interest in them is only increasing (see reviews \cite{BAR04,BAR10,BAR11a,BAR17}). Let us consider neutrinoless decay:

\begin{equation}
(A,Z) \rightarrow (A,Z-2) + 2e^{+} 
\end{equation}
\begin{equation}
e^- + (A,Z) \rightarrow (A,Z-2) + e^{+} + X
\end{equation}
\begin{equation}
e^- + e^- + (A,Z) \rightarrow (A,Z-2)^* \rightarrow (A,Z-2)+\gamma + 2X
\end{equation}

The existence of these processes means that similarly to the $0\nu\beta\beta$ decay, the lepton number is violated and the neutrino is a Majorana particle, which would require particle physics beyond the Standard Model.

Process (1) has a very nice signature. This is due to the fact that in addition to two positrons there are also
four annihilation 511 keV gamma quanta, which could be detected. But the
probability for such a search is much lower compared to $0\nu\beta\beta$ decay because
of significantly lower kinetic energy realized in such a transition and because of the Coulomb barrier for positrons.
There are only 6 candidates for such a decay: $^{78}$Kr, $^{96}$Ru, $^{106}$Cd, $^{124}$Xe, $^{130}$Ba and $^{136}$Ce (with energy of $2\beta^+$ transition $\sim$ 0.4-0.8 MeV). For most prospective isotopes their half-lives are estimated as $\sim 10^{27} - 10^{29}$ yr (for $<m_\nu> = 1$ eV) \cite{HIR94,BAR13} and this is approximately 
$10^3 - 10^5$ times longer than for the $0\nu\beta\beta$ decay for nuclei such as $^{76}$Ge, $^{136}$Xe,$^{100}$Mo, $^{82}$Se and $^{130}$Te (see reviews \cite{BAR18,BAR19,DOL19}). The best present limits for 
$2\beta^+(0\nu)$ transition are on the level $\sim 10^{21}$ yr (see reviews \cite{BAR04,BAR10,BAR11a}.

The process (2) has a good signature too and is not as strongly 
suppressed as $2\beta^+$ decay. The most optimistic estimates for the half-life give values 
$\sim 10^{26} - 10^{27}$ yr (for $<m_\nu> = 1$ eV) \cite{HIR94,BAR13}. The best present experimental limits for $\beta^+$EC$(0\nu)$ transition are on the level $\sim 10^{21}-10^{22}$ yr (see reviews \cite{BAR04,BAR10,BAR11a} and recent paper \cite{RUK18}).

In process (3) the atom de-excites emitting two X-rays and the
nucleus de-excites emitting one bremsstrahlung photon\footnote{In \cite{DOI93} (see also discussion in \cite{SUJ04}), it was mentioned that processes with emission of inner conversion electron, $e^+e^-$ pair or two $\gamma$ are also possible. And this is especially important in the case of ECEC($0\nu$) transition with
capture of two electrons from $K$ shell (in this case the transition with emission of one $\gamma$ is forbidden \cite{DOI93}).}. 
In the case of a transition to an excited state of a daughter nuclei, besides a bremsstrahlung photon,
a one or a few $\gamma$-rays are emitted from the excited state.
The probability of the process does not depend on decay energy
and increases with decreasing bremsstrahlung photon energy and increasing Z \cite{SUJ04,VER83}. The probability of such a process, even for heavy nuclei, is low, leading to $T_{1/2} \sim 10^{28} - 10^{31}$ yr for $<m_\nu> = 1$ eV \cite{SUJ04}. The best present limits for ECEC(0$\nu$) transitions are on the order $\sim 10^{21}- 10^{22}$ yr (see reviews \cite{BAR04,BAR10,BAR11a} and recent papers \cite{RUK18,AGO16,ANG16}).

In \cite{WIN55}, it was first noted that in the case of ECEC($0\nu$) a resonant condition can be realized for transition to an excited level in the
daughter nucleus with a "correct energy" (when decay energy to this level is close to zero). The same idea was proposed for transition to the ground state in 1982 \cite{VOL82}. One year later the $^{112}$Sn - $^{112}$Cd (0$^+$; 1871 keV) transition was discussed \cite{BER83}. Then the idea was reanalyzed in 2004 \cite{SUJ04}. The enhancement of the transition rate was estimated as 
$\sim 10^6 - 10^{10}$ \cite{BER83,SUJ04,KRI11,ELI11}. It means that this process starts to be competitive with $0\nu\beta\beta$ decay (in the sense of the sensitivity to neutrino mass).
There are many candidates for such resonant transitions, to the ground ( $^{152}$Gd, $^{164}$Eu and $^{180}$W) as well as to the excited states of the daughter nuclei ( $^{74}$Se, $^{78}$Kr, $^{96}$Ru, $^{106}$Cd, $^{112}$Sn, $^{124}$Xe, $^{130}$Ba, $^{136}$Ce, $^{144}$Sm, $^{152}$Gd, $^{156}$Dy, $^{162}$Er, 
$^{168}$Yb, $^{180}$W, $^{184}$Os and $^{190}$Pt) (see \cite{KRI11}, for example). The accuracy of matching to the resonance ought to be
better than 1 keV. Thus, to choose the best candidate from the above list it is necessary to know the atomic mass difference with an accuracy much better than 1 keV. The measurements have been done for all above mentioned isotopes, and in a few cases resonance conditions were found. But in all these cases ($^{152}$Gd \cite{ELI11a}, $^{106}$Cd \cite{GON11}, $^{156}$Dy \cite{ELI12} and $^{190}$Pt \cite{EIB16}) there is an additional suppression of the decay rate due to "not optimum" quantum numbers of the corresponding excited states (for example, $2^+$ , $2^-$ , $1^-$ ,...) and "not optimum" orbits of atomic electrons involved in the process (for example, $\it LL, NM, KN,$...). And the most optimistic predictions for $T_{1/2}$ are in the order of $\sim 10^{27} -10^{30}$ yr only (for $<m_\nu> = 1$ eV) \cite{KOT14,ROD12}. Thus, such experiments cannot be competitive with experiments for search for $0\nu\beta\beta$ decay. 
The best present experimental limits for possible  resonant transitions are on the level $\sim 10^{21}$ yr ($^{112}$Sn \cite{BAR11}, $^{106}$Cd \cite{BEL14} and $^{78}$Kr \cite{GAV13}). At the same time, it should be noted that there is an unsatisfactory knowledge about the excited states, and there is still a chance that "promising" candidates can be found. 

A search for $^{74}$Se- $^{74}$Ge (1204.2 keV) resonant transition was first time proposed in \cite{FRE05}. The decay scheme of this transition is presented in Fig. 1. The atomic mass difference $\Delta M$ ( $^{74}$Se and $^{74}$Ge) was known at that time with accuracy $\pm$ 2.3 keV (one standard deviation), and it was possible talk about a possible resonant capture of two electrons from $\it L$ shell for the transition to the 1204.2 keV level in $^{74}$Ge. This process would be accompanied by a cascade of two $\gamma$-quanta with energies 608.4 and 595.8 keV (68.5\%) or one $\gamma$-quantum with energy 1204.2 keV (31.5\%). 
For the first time the search for this transition was done in \cite{BAR07}. But later  atomic mass difference $\Delta M$ was measured with high accuracy (Q = 1209.169(49) keV \cite{KOL10} and Q = 1209.240(7) \cite{MOU10}) and it was demonstrated that the resonance condition is not met in this case. Nevertheless, such measurements were done again in \cite{FRE11,JEV15} and \cite{LEH16}. In \cite{LEH16} it was shown that limits in \cite{FRE11} and \cite{JEV15} were overestimate in $\sim$ 25-1000 times because of incorrect analysis of the obtained data. So, we will consider  here only the results of Ref. \cite{BAR07} and Ref. \cite{LEH16}.

Finally, let us look at the two neutrino modes of $2\beta^+$, $\beta^+$EC and ECEC
processes:

\begin{equation}
(A,Z) \rightarrow (A,Z-2) + 2e^{+} +2\nu
\end{equation}

\begin{equation}
e^- + (A,Z) \rightarrow (A,Z-2) + e^{+} + 2\nu + X
\end{equation}

\begin{equation}
e^- + e^- + (A,Z)  \rightarrow (A,Z-2)+ 2\nu + 2X
\end{equation}

These processes are not forbidden by any conservation laws. Processes (4) and (5) are strongly suppressed
because of low phase-space volume. From the experimental point of view it is very difficult to investigate the process (6), because one has to detect only low energy X-rays (or Auger electrons). Nevertheless, exactly the process (6) was detected in a few nuclei. In geochemical experiment with $^{130}$Ba ECEC($2\nu$) this process was detected with half-life value $(2.2 \pm 0.5)\times10^{21}$ yr \cite{MES01}. Recently, positive results from direct counting-rate experiments were
reported for $2K(2\nu)$ process in $^{78}$Kr \cite{RAT17} and $^{124}$Xe \cite{APR19} with half-life $(1.9^{+1.3}_{-0.8})\times10^{22}$ yr and $(1.8 \pm 0.5)\times10^{22}$ yr, respectively (see discussion about these results in \cite{BAR19a}).

This work is devoted to search for $\beta^+$EC and
ECEC processes in $^{74}$Se.

\begin{figure*}
\begin{center}
\resizebox{0.75\textwidth}{!}{\includegraphics{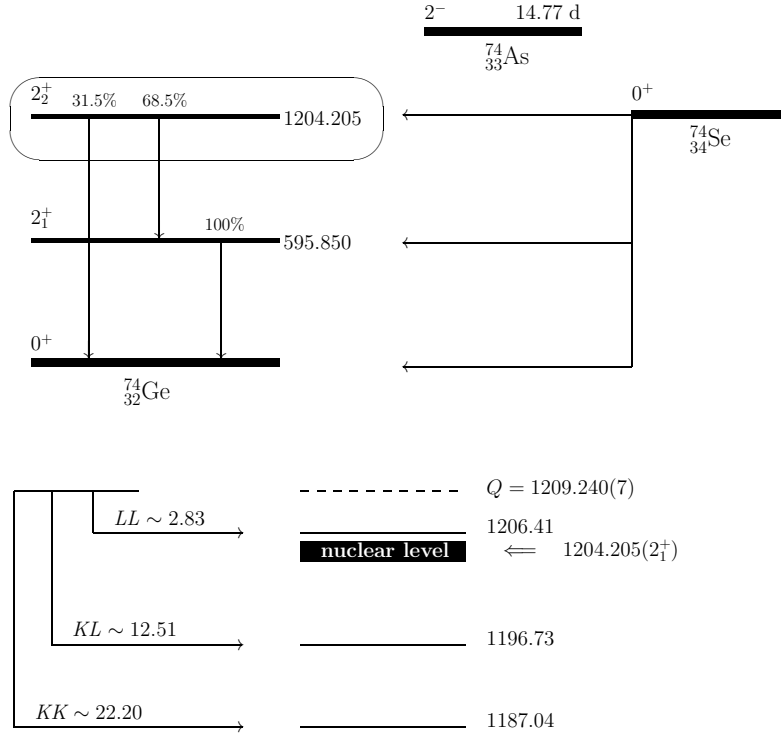}}
\caption{Energetics of the $^{74}$Se ECEC$0\nu$ decay indicating the near degeneracy of the $^{74}$Se
ground state and the second excited state in $^{74}$Ge. The circle marks the part, which is magnified 
in the lower part of the figure. Energy is indicated in keV.}  
\label{fig_1}
\end{center}
\end{figure*}

\section{Experimental}

The experimental work has been performed in the Modane Underground Laboratory (LSM, France, 4800 m w.e.), which provides the suppression of a muon flux by $\sim 2\times10^6$ times and fast neutrons 
by $\sim 10^3$ times. The natural selenium powder sample was measured using the detector OBELIX \cite{BRU17}.

The low background HPGe detector, OBELIX, was produced by the company Canberra. The detector is 
a P-type crystal with a sensitive volume of 600 cm$^3$. The mass of the detector is approximately 3.2 kg and the detector relative efficiency is 160\%. The crystal was mounted in an ultra low background U-type cryostat. The energy resolution of the detector is $\sim$ 1.2 keV (FWHM) at 122 keV ($^{57}$Co) and $\sim$ 2 keV (FWHM) at 1332 keV ($^{60}$Co).

\begin{figure*}
\begin{center}
\resizebox{0.75\textwidth}{!}{\includegraphics{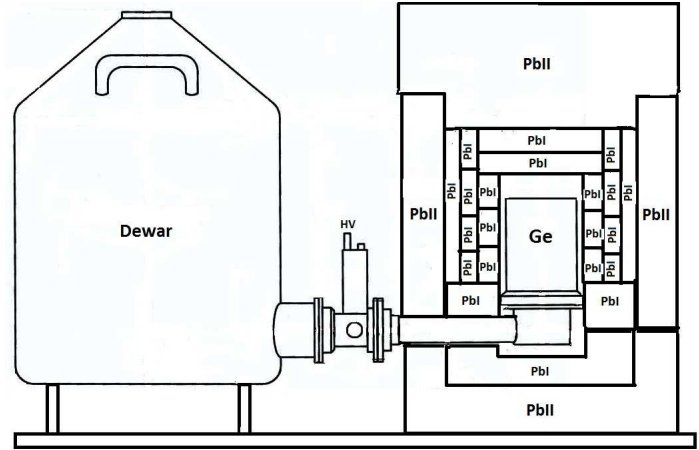}}
\caption{HPGe detector in a passive shielding.}  
\label{fig_2}
\end{center}
\end{figure*}

The detector part of the cryostat is encircled by passive shielding of several layers of lead. The Roman
lead (PbI) with a total thickness of $\sim$ 12 cm (activity of $<$ 60 mBq/kg) and 
low-activity (PbII) lead (activity of $\sim$ 10-20 Bq/kg \cite{LOA11}) with a total thickness of $\sim$ 20 cm, and placed inside a tightly closed stainless steel cover (see Fig. 2).
The electronic part of the spectrometer is based on NIM electronic modules produced by Canberra. A 3106D High Voltage module of Canberra supplies a voltage of +6000 V to the detector. Signals from the preamplifier PSC761 coupled to the detector part are passed through the Spectroscopy Amplifier 2022 and then digitized by 16384 channels ADC Multiport II. The energy threshold of the HPGe detector is about 10 keV. The software used for data taking and analysis of spectra is Genie 2000 version 3.2.1. 

A detailed description of the detector and its characteristics can be found in \cite{BRU17}.

The sample of natural selenium powder was packed in a plastic box and put on the endcap of the HPGe detector. The mass of the powder was 1600 g and mass of $^{74}$Se - 14.24 g (natural abundance is 0.89\%). The measurement time was 3283.45 hours.

A search for different  $\beta^+$EC and ECEC processes in $^{74}$Se has been performed using the
HPGe detector to look for $\gamma$-ray lines corresponding to these processes. $Q'$ is the 
effective $Q$-value defined as $Q'=\Delta M - \epsilon_1-\epsilon_2$ where 
 $\Delta M$ is the difference of parent and daughter atomic masses, $\epsilon_i$ are
electron binding energy in the daughter nuclide.

The ECEC$(0\nu)$ transitions were considered for three cases of electron captures as it is shown in Fig.~1.\\
1) Two electrons are captured from the $\it L$-shell. $Q'$ is $\sim 1206.41$ keV and
three transitions are investigated, i.e. \\
a) to the second $2^+$ level of $^{74}$Ge (1204.20 keV), accompanied by 
  de-excitation $\gamma$-quanta, 595.85 and 608.35 keV (68.5\%) or 1204.20 keV (31.5\%); \\
b) to the first $2^+$ level of $^{74}$Ge (595.85 keV),  
accompanied by the bremsstrahlung $\gamma$-quantum (610.56 keV) 
  and one de-excitation $\gamma$-quantum (595.85 keV); \\
c) to the ground state of $^{74}$Ge,  
accompanied by the bremsstrahlung $\gamma$-quantum (1206.41 keV). \\
2) One electron is captured from the $\it K$-shell, another from the $\it L$-shell.
$Q'$ is $\sim 1196.73$ keV and
two transitions are investigated, i.e. \\
a) to the first $2^+$ level of $^{74}$Ge,  
accompanied by the bremsstrahlung $\gamma$-quantum (600.88 keV) 
  and one de-excitation $\gamma$-quantum (595.85 keV); \\
b) to the ground state of $^{74}$Ge,  
accompanied by the bremsstrahlung $\gamma$-quantum (1196.73 keV). \\
3) Two electrons are captured from the $\it K$-shell.
 In this case $Q'$ is $\sim 1187.04$ keV and
two transitions are investigated, i.e. \\
a) to the first $2^+$ level of $^{74}$Ge,  
accompanied by the bremsstrahlung $\gamma$-quantum (591.19 keV) 
  and one de-excitation $\gamma$-quantum (595.85 keV); \\
b) to the ground state of $^{74}$Ge,  
accompanied by the bremsstrahlung $\gamma$-quantum (1187.04 keV). 

 The  $\beta^+$EC$(0\nu + 2\nu)$ transition is possible only to the ground state of $^{74}$Ge, accompanied by one positron which gives two  annihilation $\gamma$-quanta.

The ECEC$(2\nu)$ transitions, accompanied by detectable $\gamma$-rays, are  \\
a) the transition to
 the second $2^+$ level of $^{74}$Ge with the $\gamma$-rays, 
595.85 and 608.35 keV (68.5\%) or 1204.20 keV (31.5\%), \\
b) the transition to the first $2^+$ level of $^{74}$Ge with 
    one de-excitation $\gamma$-quantum (595.85 keV). \\

\begin{figure*}
\begin{center}
\resizebox{0.75\textwidth}{!}{\includegraphics{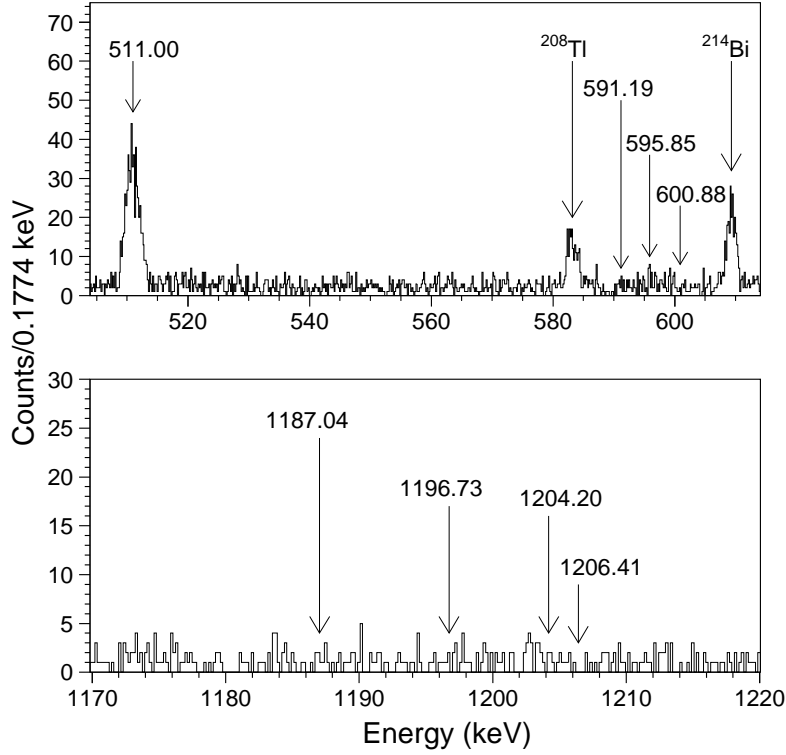}}
\caption{Energy spectra with 1600 g of natural Se in the ranges of investigated $\gamma$-rays. The measurement time is 3283.45 hours.}  
\label{fig_3}
\end{center}
\end{figure*}

The $\gamma$-ray spectra in the energy ranges corresponding to the different decay modes of $^{74}$Se 
are shown in Fig.~3. 
No extra events (statistically significant, i.e. more than $3\sigma$ over background) are observed for investigated energies. 

The limits on transitions of $^{74}$Se to the ground and 
excited states of $^{74}$Ge were estimated according to the procedure of
 the Particle Data Group \cite{TAN18} using 
the Bayesian approach (section
 39.4.1). The every bin of the spectrum is supposed to have a 
Poisson
 distribution with its mean $\mu_i$ and the number of events equal to the
 content of this $i$th bin.
The mean $\mu_i$ can be written in general form
 as
\begin{equation}
\mu_i = N\sum_{m} {\varepsilon_m a_{mi}} +
 \sum_{k}
{P_k a_{ki}} + b_i
\end{equation}
The first term describes the
 contribution of the investigated process that may have a few $\gamma$-lines
contributing appreciably to the $i$th bin. In (7) the parameter $N$ is the
 number of decays,
$\varepsilon_m$ is the detection efficiency of the $m$th
 $\gamma$-line of the transition under study and $a_{mi}$ is the
part of $m$th line
 covering the $i$th bin. For low-background measurements a $\gamma$-line may
 be taken
in a gaussian shape. The second term gives contributions of
 background $\gamma$-lines. 
Here $P_k$ is the area of the $k$th
 $\gamma$-line and $a_{ki}$ is its part covering 
the $i$th bin. The third
 term represents so named "continuous background" $b_i$ obtained by a
 spectrum smoothing
after rejecting all peaks.
The likelihood function is
 the product of probabilities for bins selected for the investigated
 process. 
Normalizing it to 1 on parameter $N$ it 
becomes probability
 density function for $N$  which is used to calculate
limits for $N$.

Limits have been calculated for different combinations of $\gamma$ -lines corresponding to the transitions under study. The best results are given in Table 1. For transitions to the  $2^+_2$  level, the limits are given for the joint analysis of two gamma quanta (595.85 and 1204.20 keV). Taking into account the gamma line of 608.35 keV led to a worse value for the limit, since this line coincides with the intense line from $^{214}$Bi (609.32 keV) and this limit is not presented. For ECEC $(0\nu)$ transitions to the $2^+_1$ level, the limits for individual $\gamma$-quanta are given, since the combined consideration of both gamma quanta does not lead to a more stringent limit. The limit for the $\beta^+$EC$(0\nu + 2\nu)$ transition was determined from the 511 keV line. It was conservatively assumed that all recorded events belong to this transition.


\begin{table}[h!]
\label{Table1}
\caption{The limits on double beta decays of $^{74}$Se. The second column presents gamma-rays in keV 
and their efficiencies used to estimate half-lives. Limits on half-lives $T_{1/2}$ are given at 
90\% C.L.}
\begin{center}
\begin{tabular}{lllll}
\hline
Transitions &  $\gamma$-ray (efficiency) & & $T_{1/2}, 10^{19}$ yr \\
to $^{74}$Ge & & This work & \cite{BAR07} & \cite{LEH16}\\
\hline
ECEC$(0\nu); LL$  &  595.85 keV (1.23\%) \\ & + & 1.10 & 0.55 & 0.70 \\
to $2^+_2$(1204.20-keV) & 1204.20 keV (0.57\%)    \\
\hline
ECEC$(0\nu); LL$   & 595.85 keV (1.82\%) & 1.58 & 1.30 & 0.82 \\
to $2^+_1$(595.85 keV)  \\                                
\hline
ECEC$(0\nu); LL$ &  1206.41 keV (1.67\%) & 6.47 & 0.41 & 0.58 \\
to g.s.  \\
\hline
ECEC$(0\nu); KL$ & 600.88 keV (1.81\%) & 4.37 \\ & & & 1.12 & 0.82\\
to $2^+_1$(595.85 keV) & 595.85 keV (1.81\%) & 1.57 \\
\hline
ECEC$(0\nu); KL$  &  1196.73 keV (1.67\%) & 3.48 & 0.64 & 0.96 \\
to g.s.  \\
\hline
ECEC$(0\nu); KK$ & 591.19 keV (1.81\%) & 4.39 \\ & & & 1.57 & 1.43  \\
to $2^+_1$(595.85 keV) & 595.85 keV (1.81\%) & 1.57  \\
\hline
ECEC$(0\nu); KK$ &  1187.04 keV (1.67\%) & 4.83 & 0.62 & -\\
to g.s.  \\
\hline
ECEC$(2\nu)$ & 595.85 keV (1.23\%) \\ & + & 1.10 & 0.55 & 0.70 \\
to ($2^+_2)$(1204.20 keV) & 1204.20 keV (0.57\%)  \\
\hline
ECEC$(2\nu)$ &  595.85 keV (2.11\%) & 1.83 & 0.77 & 0.92 \\
to the $2^+_1$(595.85 keV) \\
\hline
$\beta^+$EC$(0\nu + 2\nu)$  &  511.00 keV (4.32\%) & 0.23 & 0.19 & - \\ 
to g.s.  \\
\hline
\end{tabular}
\end{center}
\end{table}


\section{Discussion}
The obtained limits on ECEC$(0\nu+2\nu)$ transitions of $^{74}$Se to excited states of $^{74}$Ge are 1.2-11.2 times higher than the limits obtained in previous experiments \cite{BAR07,LEH16}. In particular, for the potentially resonant transition into the 1204.2 keV excited state of $^{74}$Ge a lower half-life limit of $1.1\times10^{19}$ has been obtained, which is 1.6 times better than in \cite{LEH16}. In fact, the "sensitivity" of the experiment for this transition is $\sim 2.2\times10^{21}$ years. The lower value of the obtained limit is associated with the presence of extra events ($\sim$ 2$\sigma$ effect) for the 595.85 keV line. The limit on $\beta^+$EC$(0\nu + 2\nu)$ transition to the ground state of $^{74}$Ge is improved by 20\% compared to the limit in \cite{BAR07}.


For ECEC$(0\nu)$ process, the main hopes in the past were connected with realization of resonant condition for the transition to the 1204.2 keV excited state of $^{74}$Ge. Unfortunately, recently it was demonstrated (in two independent measurements \cite{KOL10,MOU10}) that resonant condition is not met in this case. $Q'$ value for $^{74}Se(0^+) - ^{74}Ge(2^+_2)$ transition is $\sim$ 2.2 keV 
(for most promising capture of electrons from $\it L$-shell). This is why there is no  enhancement for the transition in this case. And recently the half-life value for this transition was theoretically estimated to $2\times10^{42} - 1\times10^{45}$ yr \cite{SUH12}. Such a large predicted half-life value is associated not only with the absence of resonance, but also with a very small value of the Nuclear Matrix Element for this transition and with the need to capture electrons from the L shell.

One can conclude that $^{74}$Se is not a good candidate to search for $\beta^+$EC$(2\nu)$ and ECEC$(2\nu)$
processes too and chance to detect these decays is small (even taking into account the possible 
increase in the sensitivity of such experiments in the future). Nevertheless, the obtained results are interesting 
because the present experimental limits largely exclude the existence of some unexpected (exotic) processes.

Future experimental possibilities are: if 3 kg of enriched $^{74}$Se are used then after one year of measurement with the same 
HPGe detector sensitivity of such experiment would be $\sim 10^{22}$ yr. If one  investigated 200 kg 
of enriched $^{74}$Se using such installation as LEGEND \cite{ABG17}
(where 200-1000 kg low background HPGe detectors are planned for experimental use) then for 10 years of measurement the sensitivity would increase up to  
$\sim 10^{26}$ yr. 

\section{Conclusion}
A search for $\beta^+$EC and ECEC transitions of $^{74}$Se has been performed into the 595.9 keV and 1204.2 keV excited states as well as into the ground state of $^{74}$Ge. No significant signal was detected for any of the decay modes. Lower half-life limits have been obtained which are up to a factor 10 larger than previous limits. The limit for the possible resonant decay to the 1204.2 keV state was found as $1.1\times 10^{19}$ yr (90\% C.L.), which is $\sim$ (1.6-2) times stronger than previous results \cite{BAR07,LEH16}. Apparently no resonance enhancement is visible. The realization of a resonance enhancement is anyhow strongly disfavored by precision Q-value measurements \cite{KOL10,MOU10}. It is demonstrated that in the future larger-scale experiments sensitivity to ECEC(0$\nu$) processes for such isotopes can be on the level $\sim 10^{26}$ yr.

\section*{Acknowledgment} 
We thank the staff of the Modane Underground
Laboratory
for their technical assistance in running the experiment. 
This work was performed within LEA-JOULE agreement
and IN2P3-JINR collaboration agreement No.15-93 and
partly supported by the Ministry of Education, Youth
and Sports of the Czech Republic under the Contract
Number CZ.02.1.01/0.0/0.0/ 16\_019/0000766.
Portions of this work were supported by a grant from Russian Scientific Foundation (No. 18-12-00003).

 --------------------------------------------------------------

\end{document}